\documentclass{aa}
\usepackage[varg]{txfonts}
\usepackage{mathptmx,natbib,graphicx,grffile}
\bibpunct{(}{)}{;}{a}{}{,} 
\graphicspath{{figures/}}

\begin{document}

\title{The Non-Linear Growth of the Magnetic Rayleigh-Taylor Instability} 

\author{Jack Carlyle\inst{1}
  \and Andrew Hillier\inst{2} }

\offprints{J. Carlyle, \email{jack.carlyle@esa.int}}

\institute{European~Space~Agency, ESTEC, Noordwijk, The~Netherlands
  \and College~of~Engineering,~Mathematics~and~Physical~Sciences, University~of~Exeter, United~Kingdom} 

\date{Received March 2017 / Accepted July 2017}

\abstract{This work examines the effect of the embedded magnetic field strength on the non-linear development of the  magnetic Rayleigh-Taylor Instability (RTI) (with a field-aligned interface) in an ideal gas close to the incompressible limit in three dimensions. Numerical experiments are conducted in a domain sufficiently large so as to allow the predicted critical modes to develop in a physically realistic manner. The ratio between gravity, which drives the instability in this case (as well as in several of the corresponding observations), and magnetic field strength is taken up to a ratio which accurately reflects that of observed astrophysical plasma, in order to allow comparison between the results of the simulations and the observational data which served as inspiration for this work. This study finds reduced non-linear growth of the rising bubbles of the RTI for stronger magnetic fields, and that this is directly due to the change in magnetic field strength, rather than the indirect effect of altering characteristic length scales with respect to domain size. By examining the growth of the falling spikes, the growth rate appears to be enhanced for the strongest magnetic field strengths, suggesting that rather than affecting the development of the system as a whole, increased magnetic field strengths in fact introduce an asymmetry to the system. Further investigation of this effect also revealed that the greater this asymmetry, the less efficiently the gravitational energy is released. By better understanding the under-studied regime of such a major phenomenon in astrophysics, deeper explanations for observations may be sought, and this work illustrates that the strength of magnetic fields in astrophysical plasmas influences observed RTI in subtle and complex ways.} 

\maketitle 

\section{Introduction} \label{intro}

\begin{figure*}
\centering
\includegraphics[scale=0.7]{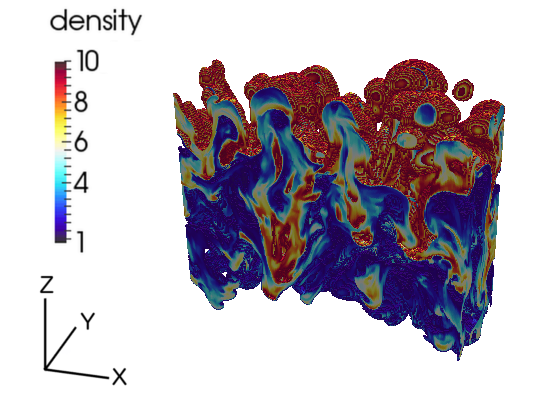}
\caption{The mixing region of Rayleigh-Taylor unstable fluids of density 1 and 10, where the pure fluids above and below the mixing region are not shown.}
\label{rtmix}
\end{figure*}

The Rayleigh-Taylor Instability (RTI) is a dynamic mixing process which occurs when a lower density fluid pushes into a higher density fluid \citep{rayleigh_investigation_1882, taylor_instability_1950}. This is usually realised by the lighter fluid supporting the heavier against gravity, and manifests as an interpenetration of finger-like plumes. The contact discontinuity between the two fluids is unstable to perturbations that grow by converting potential energy to kinetic energy, causing bubbles of the low-density fluid to rise, and spikes of the high-density fluid to sink. A thorough review of the hydrodynamic RTI is given by \cite{sharp_overview_1984}. The mixing region\footnote{In this paper, the word ``mixing'' is used to refer to the region in height of RT-unstable material that contains a combination of both the higher- and lower-density fluids. Whilst this is not necessarily a new, mixed phase fluid ({\emph i.e.}, another common definition of `mixing'), the initial regions of pure high- and low-density plasma do indeed mix together in the RTI and so in the interests of eloquence, this mass-redistribution-layer is henceforth referred to as the mixing region.} of a simulation of the RTI is shown in Figure~\ref{rtmix}.

RTI is observed to play a role in many astrophysical systems, not least of all the Sun: the observed upflows in solar prominences have been confirmed to be RT unstable using 3D numerical magnetohydrodynamic (MHD) simulations \citep{hillier_numerical_2012}; the dynamics of back-falling plasma following a failed coronal mass ejection (CME) displayed morphology indicative of RTI, which was confirmed from Alfv\'en velocity estimates \citep{innes_break_2012}; further evidence of RTI was found following the same eruption at later times by \cite{carlyle_investigating_2014}, who used evidence of the RTI to estimate magnetic field strength; RTI has been found to be a plausible mechanism for driving jets in supra arcade downflows (SADs) by \cite{guo_rayleigh-taylor_2014}; the magnetic RTI has also been investigated in relation to filamentation of emerging flux \citep{isobe_filamentary_2005}, and in the breaking up of magnetic flux sheets \citep{cattaneo_nonlinear_1988}. Further afield in the Universe, the structural formation of the crab nebula has been explained as an occurence of the RTI and a magnetic field strength was calculated by \cite{hester_wfpc2_1996}, and more recently \cite
{porth_rayleightaylor_2014} showed that the magnetic field strength in the crab nebula is not sufficient to suppress the instability, using high-resolution MHD simulations. The list goes on, with observations of the RTI reported on many scales, highlighting the importance of a thorough understanding of this process in an astrophysical context.

The magnetic RTI may be thought of as a competition between two forces: the gravitational potential, pulling the higher density fluid through the lower (and vice versa), mixing the two; and the magnetic tension, preventing deformation of the field lines and hence suppressing the mixing. Therefore it is useful to define some parameter relating these two forces so that the simulations may be compared with observations. This parameter was chosen as
\begin{equation}\label{jackpam}
J~=~\frac{c_A^2}{g\Lambda} ,
\end{equation}
where $c_A$ is the Alfv\'en velocity, $g$ is gravitational acceleration, and $\Lambda$ is a characteristic length-scale. $J$ is therefore a dimensionless parameter which describes the balance between magnetic and gravitational forces. A system with $J~\gg~1$ is likely to be dominated by the magnetic forces, and one with $J~\ll~1$ is likely to be dominated by the gravitational forces. 

Using this J parameter, it is possible to compare astrophysical plasmas observed to undergo RTI with ideal MHD simulations. For erupted filamentary solar plasma, these values are approximately $c_A~=~5~\times~10^6$~cm~s$^{-1}$ and $\Lambda~=~10^{10}$~cm from \cite{innes_break_2012}, and $g~=~10^4$~cm~s$^{-2}$ by using the surface gravity of the Sun $g_{surf}~=2.74~\times~10^4$~cm~s$^{-2}$ and taking into consideration that the plasma examined is apparently between 0.5~$-$~1~solar radii $R_{\odot}$. This gives $J~=~0.25$. Alternately, from \cite{hester_wfpc2_1996} we obtain values of approximately $c_A~=~7~\times~10^6$~cm~s$^{-1}$, $g~=3.5~\times~10^{-3}$~cm~s$^{-2}$, and $\Lambda~=~10^{17}$~cm, which gives $J~=~0.14$. The strongest magnetic fields studied in previous 3D numerical experiments appear in \cite{stone_magnetic_2007} and are conducted in the $J~=~0.03$ regime, so in order to better investigate the RTI in the context of astrophysical plasmas, a higher value of $J$ should be explored.

Much insight can be gained into the RTI from analytic work carried out by \cite{chandrasekhar_hydrodynamic_1961}. If the fluids involved in the RTI are inviscid, the growth rate of the linear phase of the instability is described by the spatial frequency $\omega$:
\begin{equation} \label{rtgrow1}
\omega^2~=~-Akg ,
\end{equation}
where $A$ is the Atwood number, defined as $(\rho_u~-~\rho_l)/(\rho_u~+~\rho_l)$, $\rho$ is density (and the subscript denotes upper and lower density fluids), $k$ is wavenumber, and $g$ is gravitational acceleration. The amplitude of disturbance to the boundary $\eta$ a linear mode at a given time $t$ in this case is defined by
\begin{equation} \label{rtamplin}
\eta(t)~=~\eta_0\textrm{exp}[(Akg)^{1/2}t] ,
\end{equation}
where $\eta_0~=~\eta(t=0)$ is the size of the initial small perturbation. 

The addition of a magnetic field $B$ parallel to the contact discontinuity (provided the fluids are sufficiently conductive) modifies this linear growth rate through the addition of magnetic tension along the field which will work to suppress high wavenumber perturbations. If the magnetic field is purely in the $x$-direction, the growth rate is given by
\begin{equation} \label{rtgrow2}
\omega^2 = -Akg + \frac{\cos^2\theta k^2 B^2}{2 \pi (\rho_u + \rho_l)},
\end{equation}
\citep{chandrasekhar_hydrodynamic_1961} where $\theta$ is the angle between $k$ and $B$. Much analytic work has been performed on this linear regime of the RTI, which has been used so frequently to attempt to explain observed astrophysical processes. 

It is useful to estimate a characteristic length scale associated with the RTI, which can be achieved with these linear equations. If equation~\ref{rtgrow2} is below 0, then $\omega$ is imaginary and the system is stable and any perturbations will produce waves in the interface. If $\omega$ is real, the system is unstable to that perturbation and the instability will give rise to the bubbles and spikes described. The most unstable wavelength of the instability is always the interchange mode, where $\theta~=~\pi/2$, because it does no work against the magnetic field. However, the most unstable wavelength of the instability for a given $\theta$, \emph{i.e.}, the characteristic lengthscale of the instability at a particular angle to the magnetic field, will be at the peak of the distribution of $\omega(k, \theta)$, that is where $\partial \omega / \partial k$ = 0. Doing this gives
\begin{equation} \label{rtgrow3}
\lambda_{u} = \frac{2 \pi}{k_u} = \frac{2 \cos^2 \theta B^2}{g(\rho_u - \rho_l)} ,
\end{equation}
(\emph{e.g.} \citealt{carlyle_investigating_2014}) where $k_u$ is the wavenumber of the most unstable mode, and $\lambda_u$ is the corresponding wavelength of this mode: the dominant growth scale of the instability.

Once the non-linear saturation of the instability has been reached, these equations will no longer describe the development of the system. There is not a definite distinction between the linear and non-linear regimes of the RTI, in fact a transitional stage between the two exists which is not easily described by simple ordinary differential equations. The argument can be made that the non-linear saturation is reached when the contact discontinuity has been deformed in the vertical direction over a distance comparable with $1/k$ \citep{fermi_taylor_1953}. Since $k~=~2\pi~/~\lambda$, qualitatively the non-linear saturation could be described as the point at which the vertical scales are comparable to the horizontal. See Section~5 of \cite{hillier_nature_2016} for a thorough description of this.

In the non-linear regime, the growth of the (hydrodynamic) RTI becomes self similar, and is described by
\begin{equation} \label{rteqnl0}
\frac{\partial h}{\partial t}~=~2(\alpha Agh)^{1/2} ,
\end{equation}
\citep{ristorcelli_rayleigh_2004, cook_mixing_2004} where $h$ is the distance from the initial interface which the mixing region has penetrated the homogenous fluid (\emph{i.e.} the height of the bubbles or the depth of the spikes), and $t$ is time. The dimensionless coefficient $\alpha$ is referred to as the non-linear growth rate, and is insensitive to initial conditions. By taking the positive roots (\emph{i.e.}, those which are physically realisable), for constant $\alpha$, $A$ and $g$, the solution to equation~\ref{rteqnl0} is
\begin{equation} \label{rteqnl1}
h~=~\alpha Agt^2~+~2(\alpha Agh_0)^{1/2}t~+~h_0 .
\end{equation}
If $t~=~0$ is chosen as the onset of non-linearity, then $h_0$ is the thickness of the mixing region at this point (the extent of the mixing region from the initial contact discontinuity) \citep{cabot_reynolds_2006}. At later times, the first term of equation~\ref{rteqnl1} dominates, and the latter two may be neglected, so we have
\begin{equation} \label{rteqnl2}
h~\approx~\alpha Agt^2 .
\end{equation}
For more in-depth analytic work into the magnetic RTI, the authors also recommend the text books by \cite{goedbloed_principles_2004} and \cite{goedbloed_advanced_2010}.

By comparing, it can be seen that the exponential growth of the linear regime will compete with the $t^2$ growth of the non-linear regime at smaller $k$, and so $h$ should be greater than the maximum wavelength of the system in order for equation~\ref{rteqnl2} to be applicable. It is also apparent that in a simulation, the characteristic domain size $L$ should be greater than the characteristic separation of the most unstable mode $\lambda_u$, and as a good rule of thumb this is assumed to be of the same order as $h$. Therefore, non-linear analysis of the magnetic RTI should be conducted in a system which satisfies $L~>~h~>~\lambda_{max}$. 

Non-linear growth rate $\alpha$ has been determined from multiple laboratory experiments to be approximately 0.057, however, studies of simulations of the RTI typically give a value a factor two smaller than this. \cite{glimm_critical_2001} conclude that numerical dissipation effects (such as mass diffusion and viscosity) due to algorithmic differences and differences in simulation duration are the main reasons for the observed spread in non-linear growth rate across studies, and \cite{dimonte_comparative_2004} argue that the reduced growth rate in simulations arises from band-limited initial perturbations.

\cite{Jun_numerical_1995} studied the linear and non-linear regimes of the RTI using 2D MHD simulations, investigating the effect of a magnetic field tangential to the initial interface as well as fields normal to this. They found enhanced growth (relative to the hydrodynamic case) in the normal case, the material collimating along field lines as the instability sets in, but there is an upper limit to the magnetic field strength, beyond which the growth is greatly suppressed. However, as pointed out by \cite{hillier_nature_2016}, unlike the hydrodynamic case (or the case of \cite{Jun_numerical_1995} where the magnetic fields are normal to the simulated plane), the evolution of the instability is no longer isotropic due to the addition of the magnetic field (tangential to the interface), and so a 2D simulation would only be able to capture the growth of a single mode from the whole spectrum of preferred modes; a fundamentally 3D system cannot be truly captured by 2D simulations.

The non-linear phase of the RTI with a magnetic field has been studied in 3D MHD simulations: \cite{stone_magnetic_2007} showed how the shape of resulting bubbles is affected by magnetic field configuration, and \cite{stone_nonlinear_2007} demonstrated that whilst the instability was slowed by the addition of a strong magnetic field during the initial onset of the instability, at later times the non-linear growth rate was actually enhanced relative to the hydrodynamic case. This is attributed to the suppression of mixing between the fluids, which occurs through secondary Kelvin-Helmholtz roll-ups at the edges of the bubbles and fingers, preserving the density discontinuity. This directly refutes the postulation above that unidirectional magnetic fields suppress the modes of the instability in one dimension, reducing the overall growth rate of the system. Beyond these investigations, little numerical work has been carried out on the non-linear saturation of the RTI. 

This study aims to conduct idealised 3D numerical MHD simulations in a parameter space approximating astrophysical plasma, and the details of these are given. It should be noted that whilst we approach realistic values for th $J$ parameter (equation~\ref{jackpam}), the simulations are still highly idealised; for example, magnetic field may not always be aligned with the interface in reality (nor indeed will the magnetic fields in each plasma be similarly oriented, or even in strength, necessarily). This work also neglects other physical processes, such as radiative transfer and ionisation balance, which are not thought to have appreciable effects on the growth of plasma instabilities under varying magnetic field. We then present analysis of the non-linear regime of the instability, particularly the growth rate $\alpha$ and the mixing of the system. Finally, we discuss the implications and validity of these results with respect to application to the observations which inspired this study.

\section{Numerical MHD Simulations} \label{meth}

This work used the Athena code for astrophysical MHD (see \cite{stone_athena:_2008} for a complete description of this code), which solves the equations of ideal MHD with a constant gravitational acceleration, $\mathbf{g}~=~(0,0,-g)$:
\begin{equation} \label{mhd1}
\frac{\partial \rho}{\partial t} + \nabla \cdotp (\rho \mathbf{v}) = 0
\end{equation}
\begin{equation} \label{mhd2}
\frac{\partial \rho \mathbf{v}}{\partial t} + \nabla \cdotp (\rho \mathbf{v} \mathbf{v} - \mathbf{BB}) + \nabla P = \rho \mathbf{g}
\end{equation}
\begin{equation} \label{mhd3}
\frac{\partial \mathbf{B}}{\partial t} + \nabla \times (\mathbf{v} \times \mathbf{B}) = 0
\end{equation}
\begin{equation} \label{mhd4}
\frac{\partial E}{\partial t} + \nabla \cdotp [(E + P) \mathbf{v} - \mathbf{B}(\mathbf{B} \cdotp \mathbf{v})] = \rho \mathbf{v} \cdotp \mathbf{g} ~.
\end{equation}
where we have used total pressure $P \equiv P_g + (\mathbf{B} \cdotp \mathbf{B}) / 2$, gas pressure $P_g = (\gamma - 1)\epsilon$, total energy density $E~\equiv~\epsilon~+~\rho(\mathbf{v}\cdotp\mathbf{v}/2~+~(\mathbf{B}\cdotp\mathbf{B})/2$, internal energy $\epsilon$, and the adiabatic index $\gamma = 5/3$. This is not the value which would necessarily be expected from \emph{e.g.} solar plasma, however, the simulations are conducted at the incompressible limit by using a large enough sound speed such that all fluid motions are highly subsonic, and so varying the adiabatic index has little effect on the results \citep{stone_nonlinear_2007}. Note that these equations have been normalised to dimensionless units such that sound speed $c_s~=~1$ (at the interface between the fluids) for $B~=~1$ and $\rho~=~\rho_u~=~1$, and the characteristic length scale of the system $\Lambda~=~1$. In this model, $g = 0.1$ and so $\sqrt{g\Lambda}/c_s~\ll~1$, which indicates that the induced flows are almost incompressible. 

The equations are solved using a second-order Godunov scheme. Perhaps the most important element of this scheme is the Riemann solver, which calculates time-averaged fluxes of all conserved quantities at cell interfaces. Here a multi-state Harten-Lax-van Leer Discontinuities (HLLD) approximate Riemann solver is used, since this is as accurate as the well studied Roe approximate Riemann solver and less computationally demanding \citep{miyoshi_multi-state_2005}. This is combined with the constrained transport (CT) technique which evolves the induction equation in a way which ensures zero divergence of the poloidal (constrained) field components to within machine round-off error \citep{evans_simulation_1988}. Discretization is based on cell-centered volume averages for mass, momentum, and energy, and face-centered area averages for the magnetic field. Athena has been shown to be successful at conducting MHD simulations of the RTI in three dimensions \citep{stone_magnetic_2007} into the non-linear regime \citep{stone_nonlinear_2007}, and as such it was deemed suitable for conducting this investigation.

The $x$ and $y$ boundaries of the domain are periodic whilst the $z$ boundaries are reflective, and the origin is in the centre of the domain. The regular cartesian grid used has dimensions of $256~\times~64~\times~1024$. Resolution in the $z$-dimension (height) was doubled relative to $x$ and $y$ -- that is, $dz~=~dx/2~=~dy/2$ -- so as to achieve a high precision and accuracy of measurement of height and therefore growth rate.

The system is initially in hydrostatic equilibrium, and the gas pressure is chosen such that the sound speed ($c_s$) in the light fluid at the interface is unity, and so
\begin{equation} \label{press}
P(z) = \frac{3}{5} - g\rho z + \frac{B^2}{2} .
\end{equation}
A characteristic length scale $\Lambda$ of roughly an order of magnitude larger than the scales predicted by equation~\ref{rtgrow3} for desired magnetic field strengths. The width of the domain in the direction of magnetic field used for the first set of simulations to be $L_x~=~0.4\Lambda$. This width is chosen to allow $L_x~\geq~\lambda_u$ for all simulations, and resolves the dominant wavelengths $\lambda_u$ with at least 44 grid points. 

RTI modes perpendicular to the magnetic field behave as the hydrodynamic case, and so the smallest scales are favoured. Numerical diffusion in the simulations was of the order $0.01\Lambda$ for the resolution used, which is measured from the extent over which the contact discontinuity reaches after the two pure fluids are allowed to relax with no perturbations given to the system. Therefore, $L_y~=~0.1\Lambda$ is used as the depth of the domain, allowing sufficient space for interchange structures to develop. A height of $L_z~=~0.8\Lambda$ is used to ensure the $L_z~\gg~h$ is not violated, and to prevent the growth of the bubbles being affected by the reflective upper boundary, as it is for this reason that \cite{stone_magnetic_2007} discarded 20\% of their data.

The lowest $J$ (corresponding to the weakest magnetic field strength, see equation~\ref{jackpam}) corresponds to the Athena RTI test case (and as such has been rigorously analysed and tested for accuracy), however, higher $J$ simulations have not previously been conducted; the highest $J$ (strongest field) used here is at the limit of $L~\simeq~\lambda_u$, violating the requirement of $L~>~h~>~\lambda_{max}$ (this was further investigated in the second set of simulations, which are described below). A larger L was not used as the simulations were already computationally demanding; a lower resolution was also avoided as the current setup should lead to approximately 50 pixels per $\lambda_u$, and lower resolution is not desirable as it is important that the simulation allows all scales dictated by the physics to develop, and not be inhibited for computational reasons. The magnetic field is initially applied uniformly along the $x$ axis, that is ($B_x~=~const.$, $B_y~=~0$, $B_z~=~0$). Seven simulations were run in this set, and are described in Table~\ref{simtab}.

\begin{table}
    \center
    \begin{tabular}{c c c c c}
    Label & $J$ & $L_x/\lambda_u$ & $\alpha$ & $\sigma$ \\
    \hline
    B1 & 0.03 & 5.7 & 0.0509 & 0.0059 \\
    B2 & 0.04 & 4.0 & 0.0382 & 0.0054 \\
    B3 & 0.05 & 2.9 & 0.0422 & 0.0069 \\
    B4 & 0.06 & 2.2 & 0.0398 & 0.0018 \\
    B5 & 0.08 & 1.8 & 0.0406 & 0.0028 \\
    B6 & 0.10 & 1.4 & 0.0366 & 0.0013 \\
    B7 & 0.12 & 1.2 & 0.0379 & 0.0013 \\
    B8 & 0.14 & 1.0 & 0.0350 &  0.0031 \\
        \hline 
    W1 & 0.03 & 5.7 & 0.0509 & 0.0059 \\
    W2 & 0.03 & 2.9 & 0.0588 & 0.0077 \\
    W3 & 0.03 & 1.4 & 0.0380 & 0.0061 \\
    W4 & 0.03 & 0.7 & 0.0378 & 0.0093 \\
    \end{tabular}
    \caption{Relevant parameter space explored of all simulations conducted, alongside measured non-linear growth rate $\alpha$ and standard deviation $\sigma$.}
    \label{simtab}
\end{table}

\begin{figure*}
\centering
\includegraphics[scale=0.25]{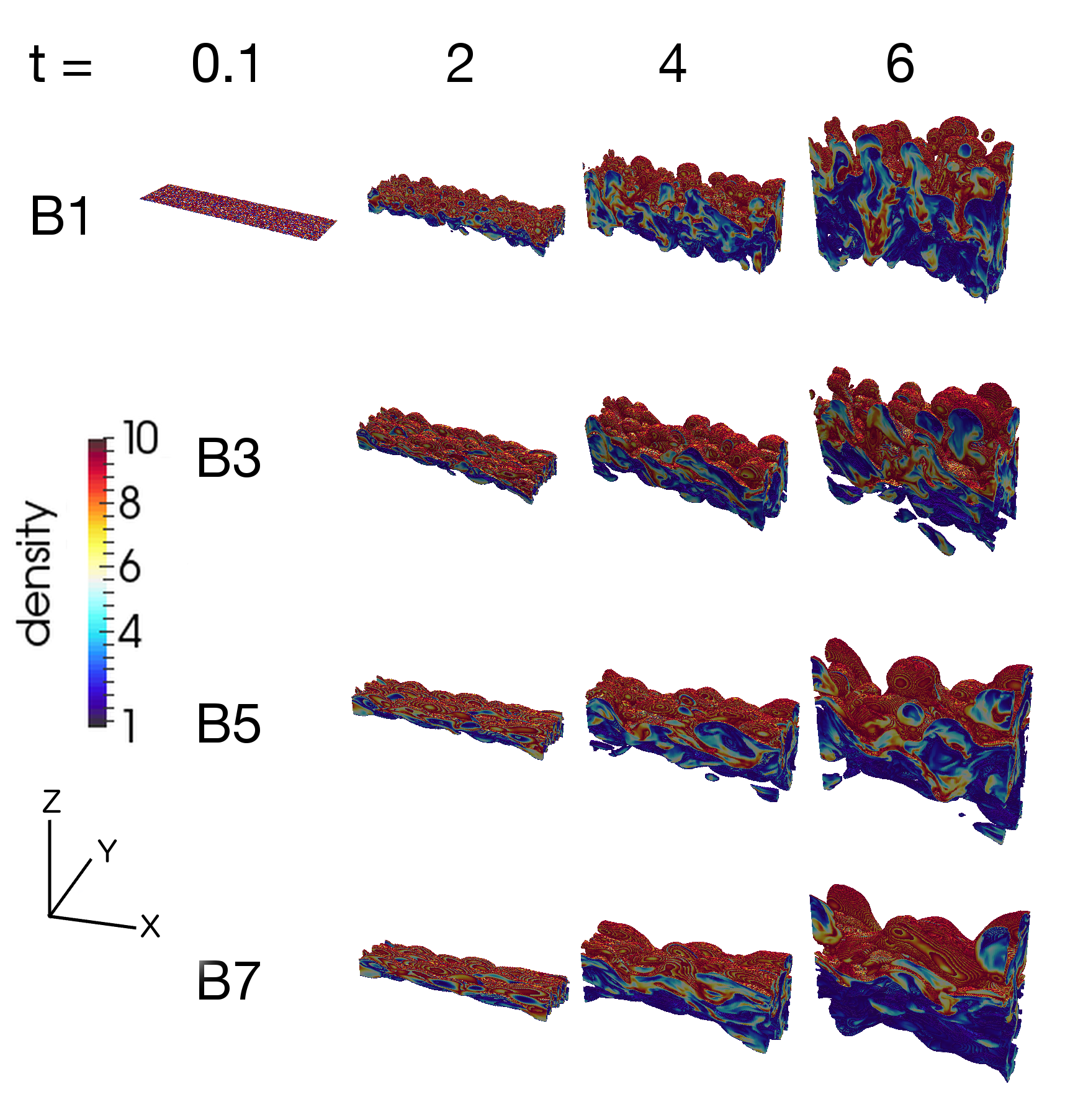}
\caption{Visualisations of the mixing region $1.5~\leq~\rho~\leq~9.5$ of simulations B1, B3, B5 and B7 (c.f. Table~\ref{simtab}) at $t~=~2,~4,~6$.}
\label{sims1}
\end{figure*}

The mixing layer (that is all fluid with $1.5~\leq~\rho~\leq~9.5$, where the initial setup has $\rho_l~=~1$ for $z~<~0$ and $\rho_u~=~10$ for $z~\geq~0$) of B1, B3, B5 and B7 are shown at three points along the run in Figure~\ref{sims1}. The chosen start time of 0.1 rather than 0 is to show the interface; at $t~=~0$, the lower half of the domain is filled with $\rho_l~=~1$ material and the upper half with $\rho_u~=~10$, so no mixing layer is visible. As the simulations progress, bubbles of scales predicted by equation~\ref{rtgrow3} can be seen developing along $x$, the direction along which $B$ is directed. The scales of these so-called undular modes are seen to increase as $J$ (and hence magnetic field strength) increases, whilst the scales across the magnetic field, the interchange mode, remain apparently constant for all simulations: one bubble along the $y$-direction can be seen.

\begin{figure*}
\centering
\includegraphics[scale=0.33,trim={2.2cm 0 0 0}]{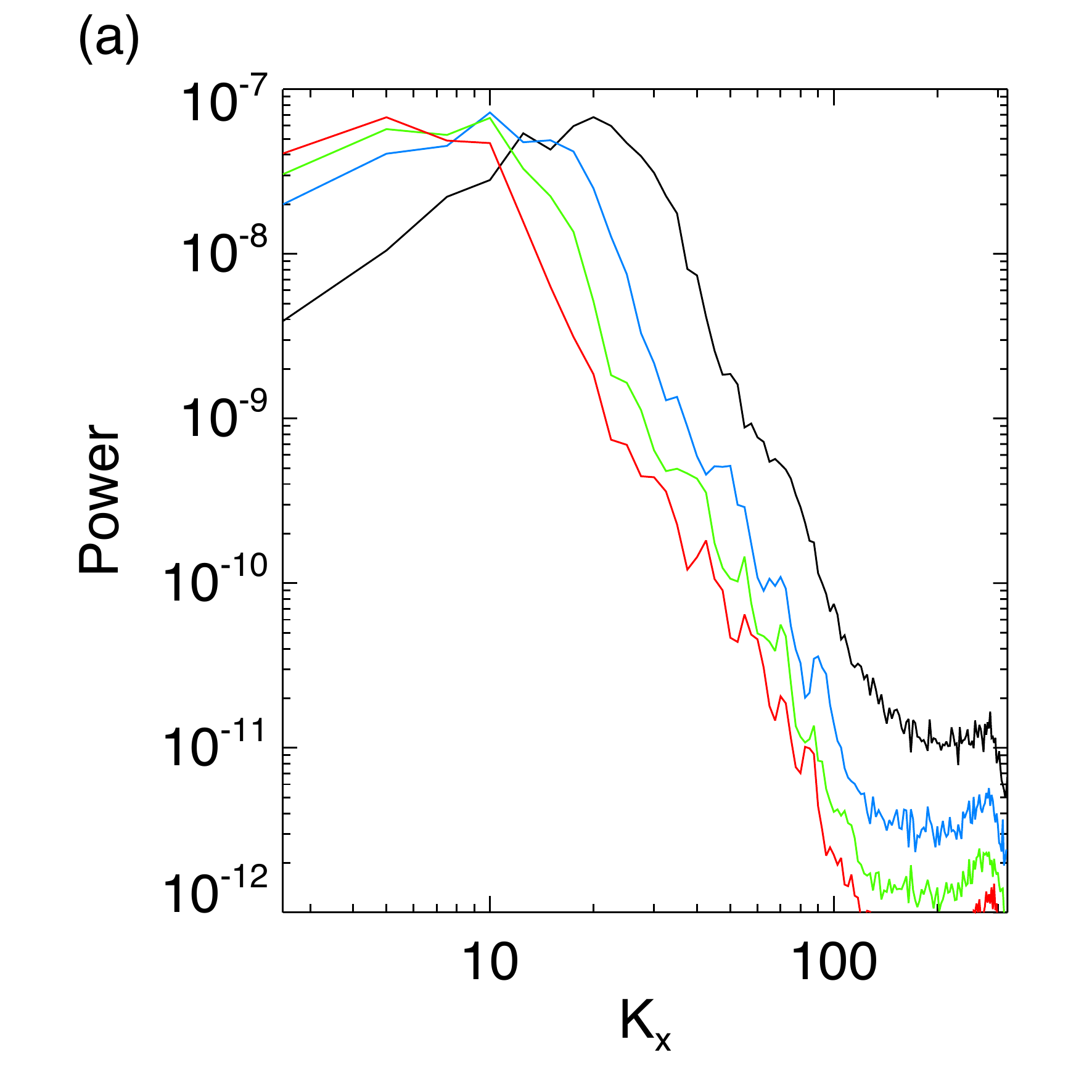}
\includegraphics[scale=0.33,trim={2.2cm 0 0 0}]{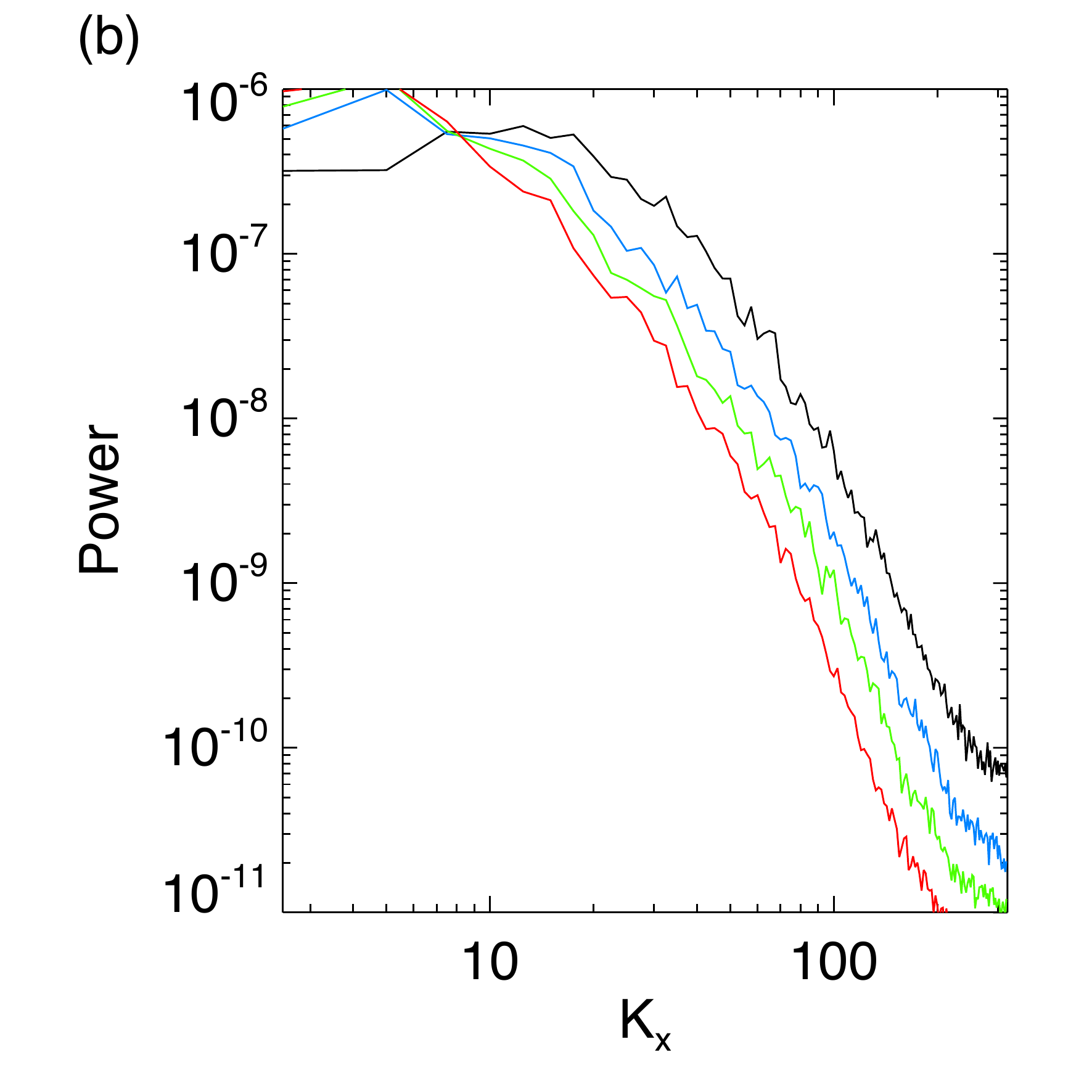}
\includegraphics[scale=0.33,trim={2.2cm 0 0 0}]{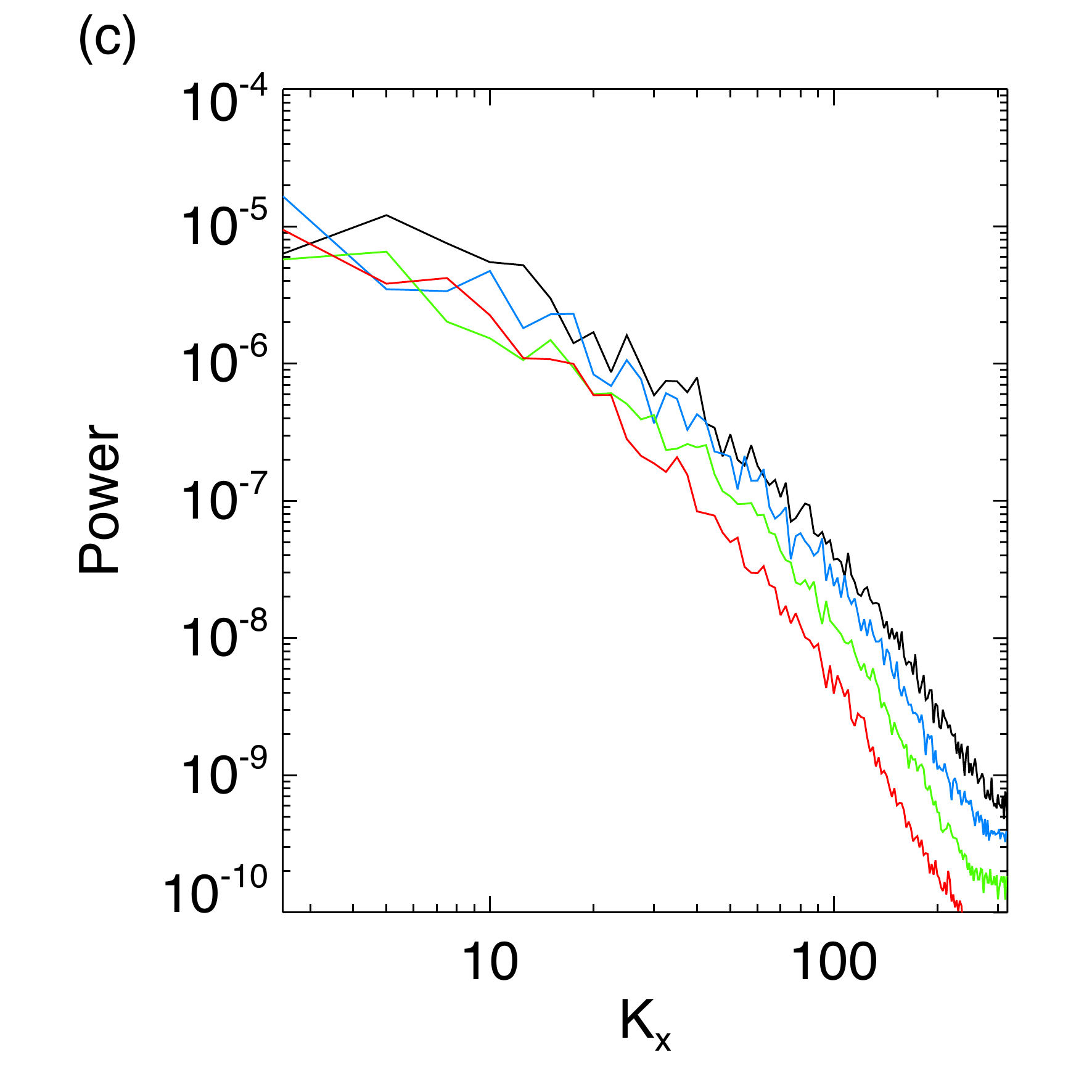}
\includegraphics[scale=0.4]{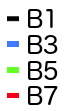} \\
\includegraphics[scale=0.33,trim={2.2cm 0 0 0}]{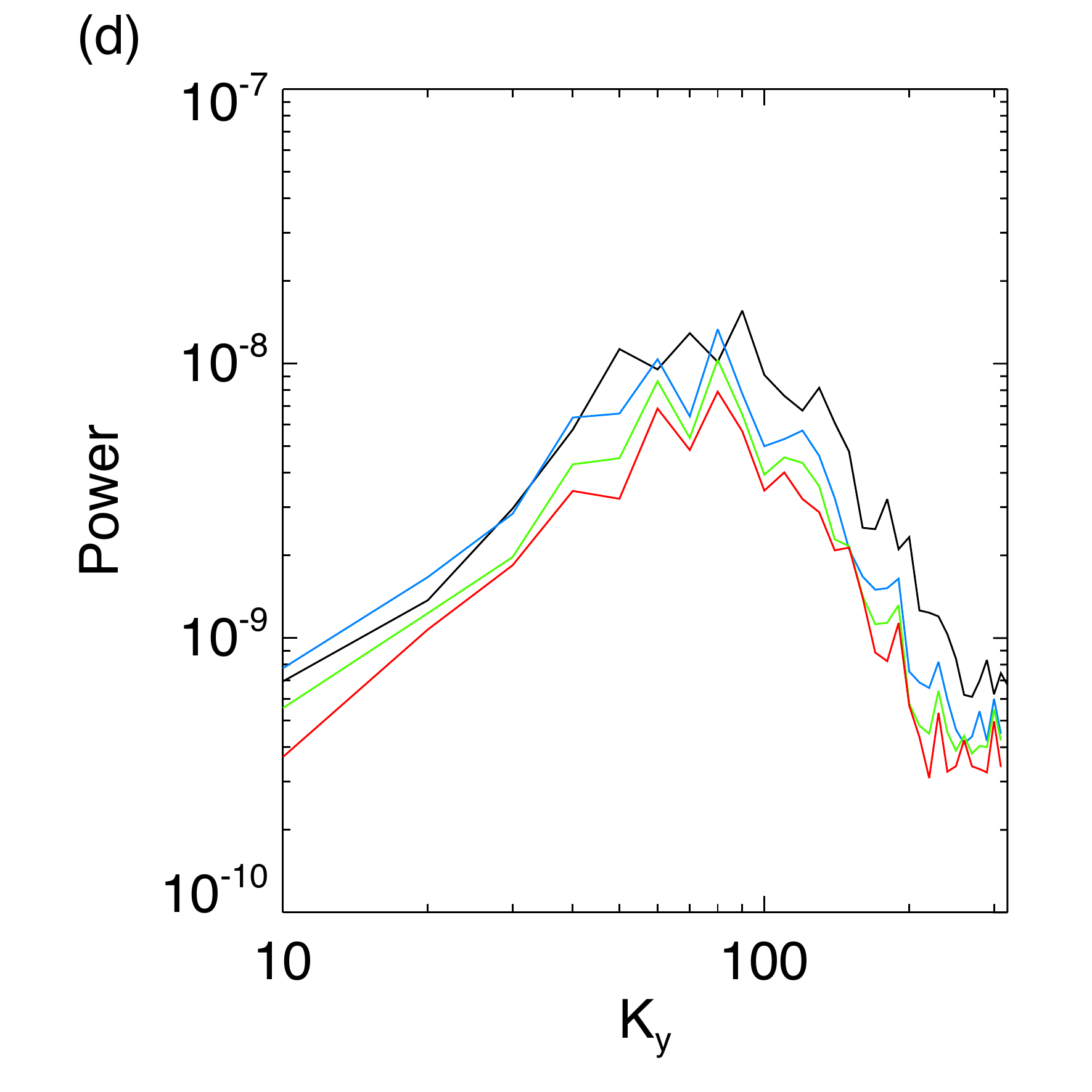}
\includegraphics[scale=0.33,trim={2.2cm 0 0 0}]{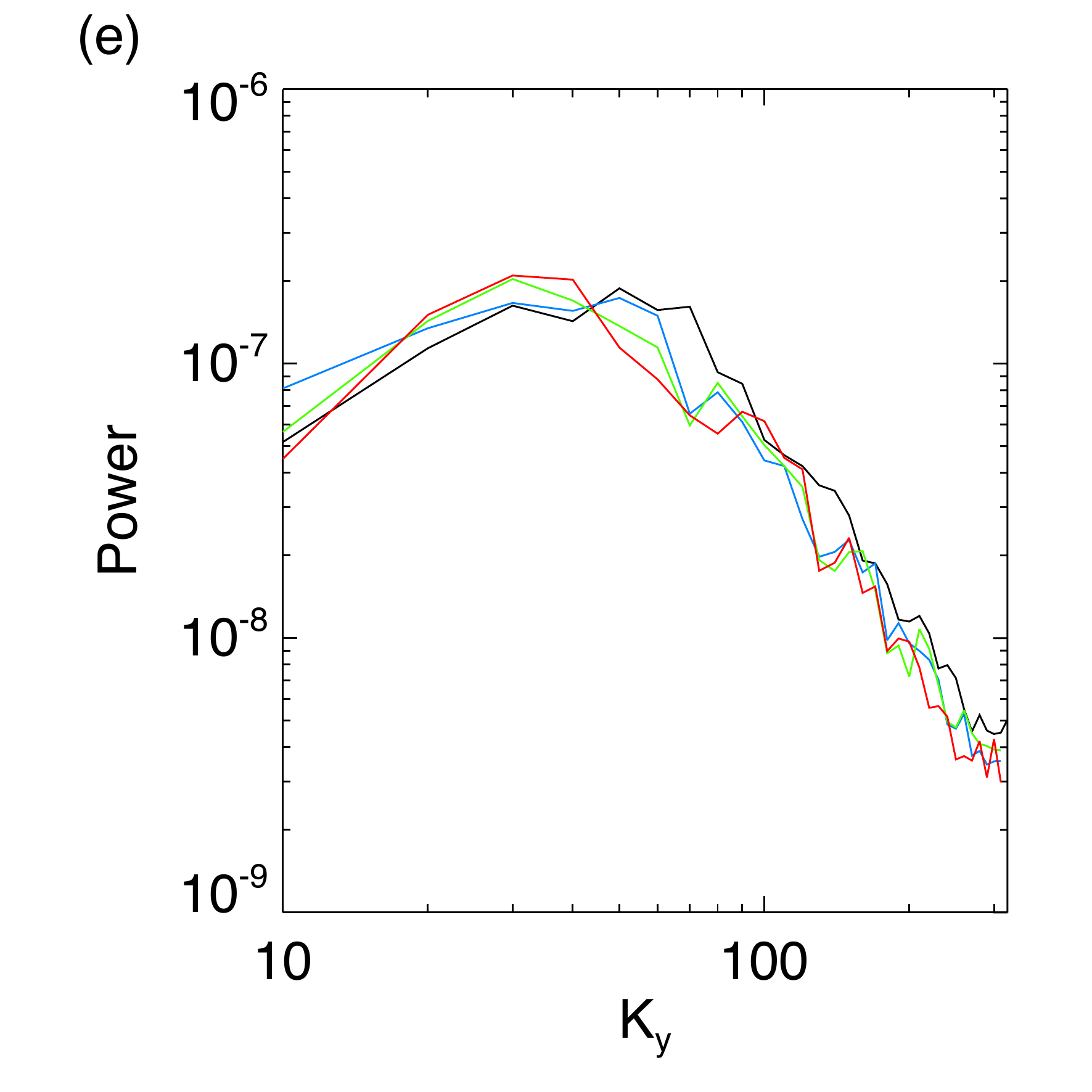}
\includegraphics[scale=0.33,trim={2.2cm 0 0 0}]{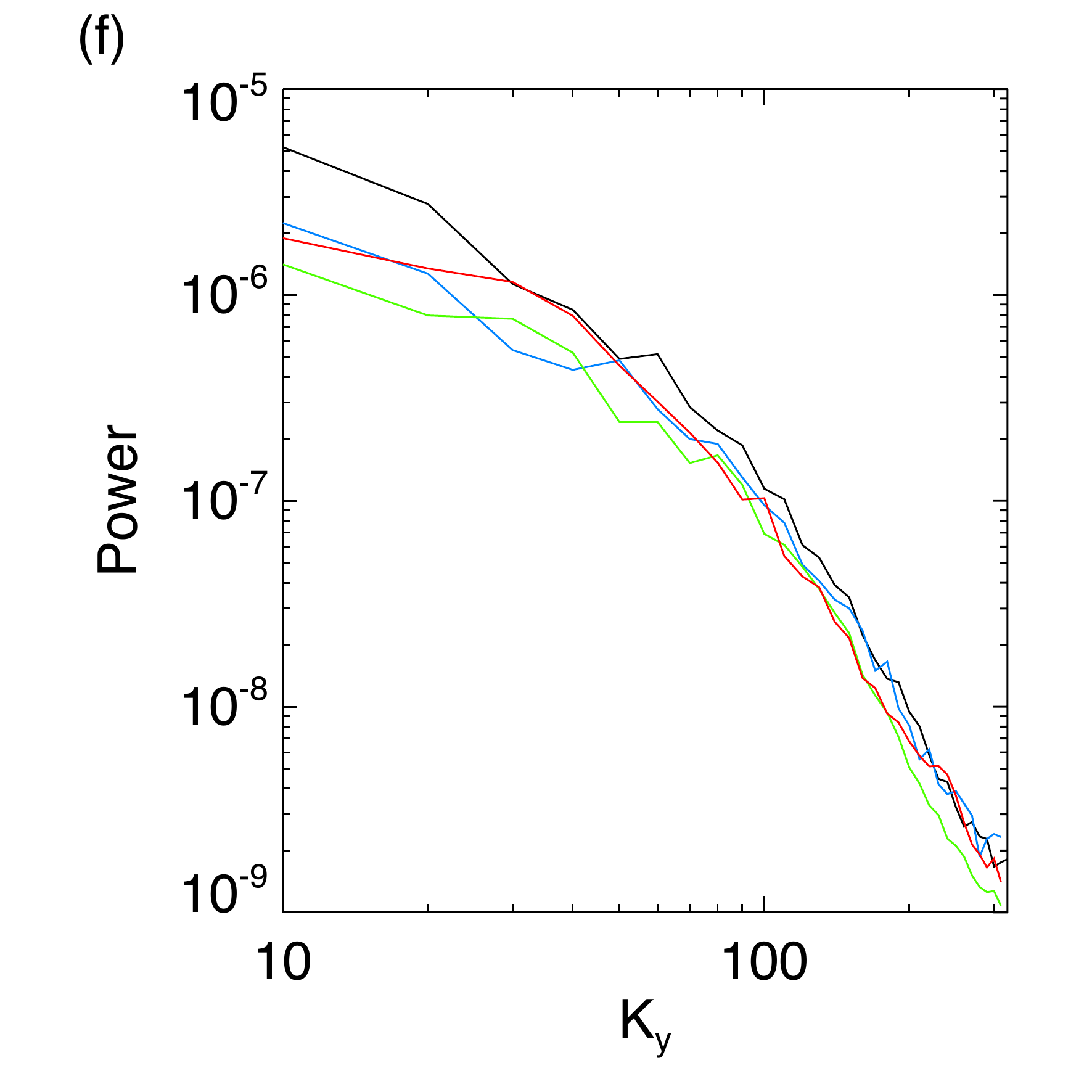}
\includegraphics[scale=0.4]{magkey2.png}
\caption{Fourier transform of $B_z$ at $t~=~1$ (a, d), $t~=~2$ (b, e), and $t~=~6$ (c, f), of the central $z$-slice for selected B simulations, showing the spectral power of different spatial frequencies; (a, b, c) shows the scales aligned with magnetic field, (d, e, f) shows scales across.}
\label{rtft}
\end{figure*}

The structures created in the linear and non-linear phases of the instability, as shown in Figure ~\ref{sims1}, are clearly dependent on the existence and strength of the magnetic field. Generally the instability drives the creation of filamentary structure that is aligned with the direction of the magnetic field. Looking at $t~=~2$ for the B3 simulation, approximately five peaks can be seen across the length of the box, but if these were just undular modes we would expect to see about three (see Table~\ref{simtab}). This implies that the formation of mixed modes, with structure across the magnetic field, is playing an important role. As the instability develops in its non-linear phase, larger structures develop both across and along the magnetic field, where the simulations with stronger magnetic fields maintain their larger aspect ratio between the along field and across field scales even into this regime. These impressions from the figure have been confirmed through a Fourier transform of the data, presented in Figure~\ref{rtft}. This shows that the power in higher frequencies is reduced for stronger magnetic field only in the direction of the magnetic field, indicating that smaller scales are suppressed by magnetic field.

\begin{figure}
\centering
\includegraphics[scale=0.15]{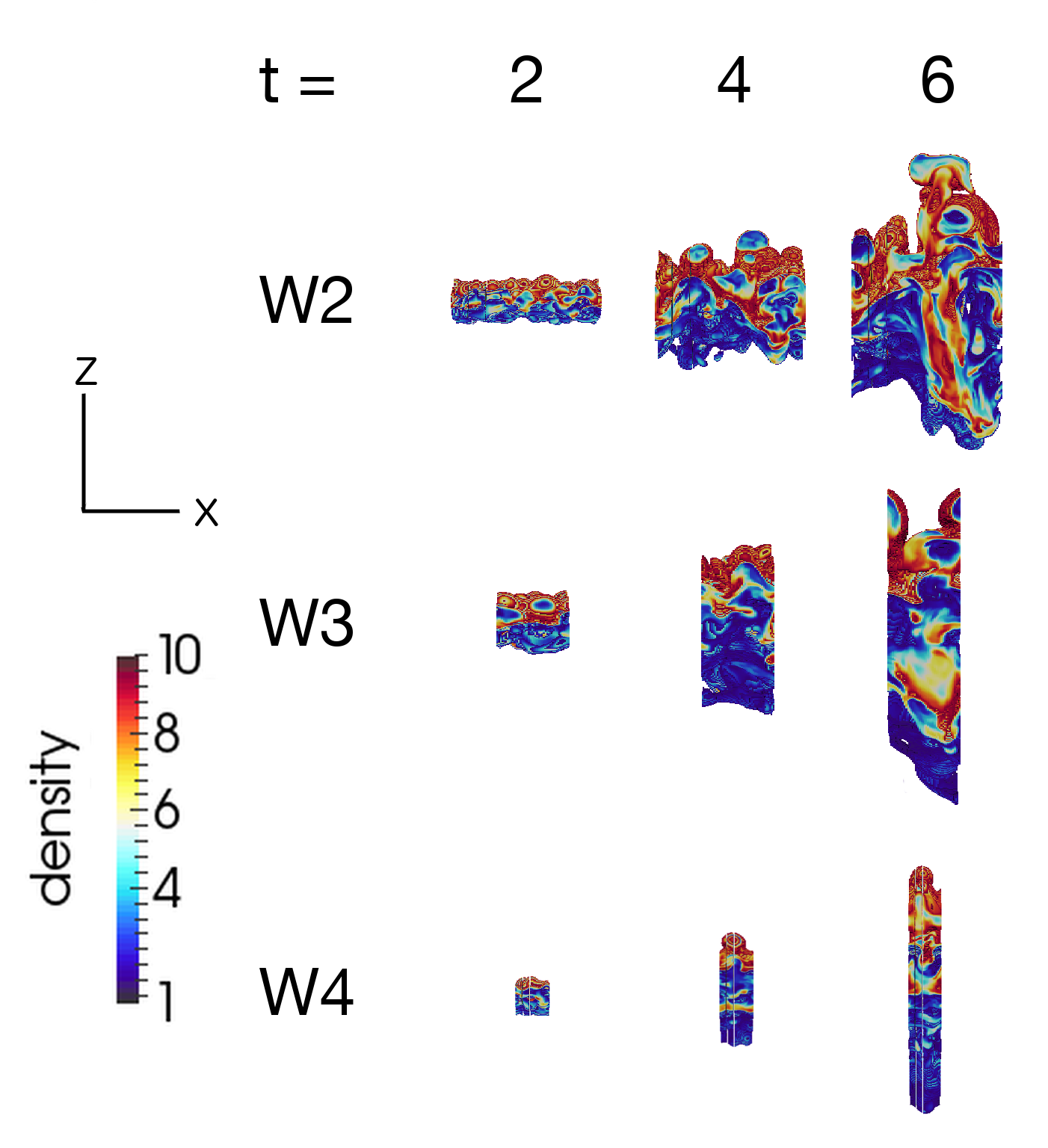}
\caption{Visualisations of the mixing region $1.5~\leq~\rho~\leq~9.5$ of simulations W2, W3 and W4 (c.f. Table~\ref{simtab}) at $t~=~2,~4,~6$.}
\label{sims2}
\end{figure}

Since the only two parameters which vary between B1 -- B7 are $J$ and $L_x/\lambda_u$ (the ratio of domain width to dominant linear wavelength of the RTI), it is necessary to ensure the observed change in growth rate is due to the former rather than the latter. In order to achieve this, a second set of simulations is run, where magnetic field is kept constant but the width of the domain is made progressively smaller. If a dependence of $\alpha$ on $L/\lambda_u$ can be seen in simulations of constant magnetic field similar to this dependence in the previous set of simulations, then the magnetic field can not be said to be causing the postulated effect on growth rate. Another benefit of this is to investigate behaviour which violates the $L_x~>~h~>~\lambda_{max}$ requirement for a constant magnetic field. Four simulations are conducted with $J~=~0.03$ and other parameters detailed in Table~\ref{simtab}. Figure~\ref{sims2} shows the mixing region for W2, 3 and 4 for $t~=~2,~4,~6$. Note that W1 has the same initial conditions as B1.

\section{Analysis} \label{res}

\begin{figure}
\centering
\includegraphics[scale=0.55]{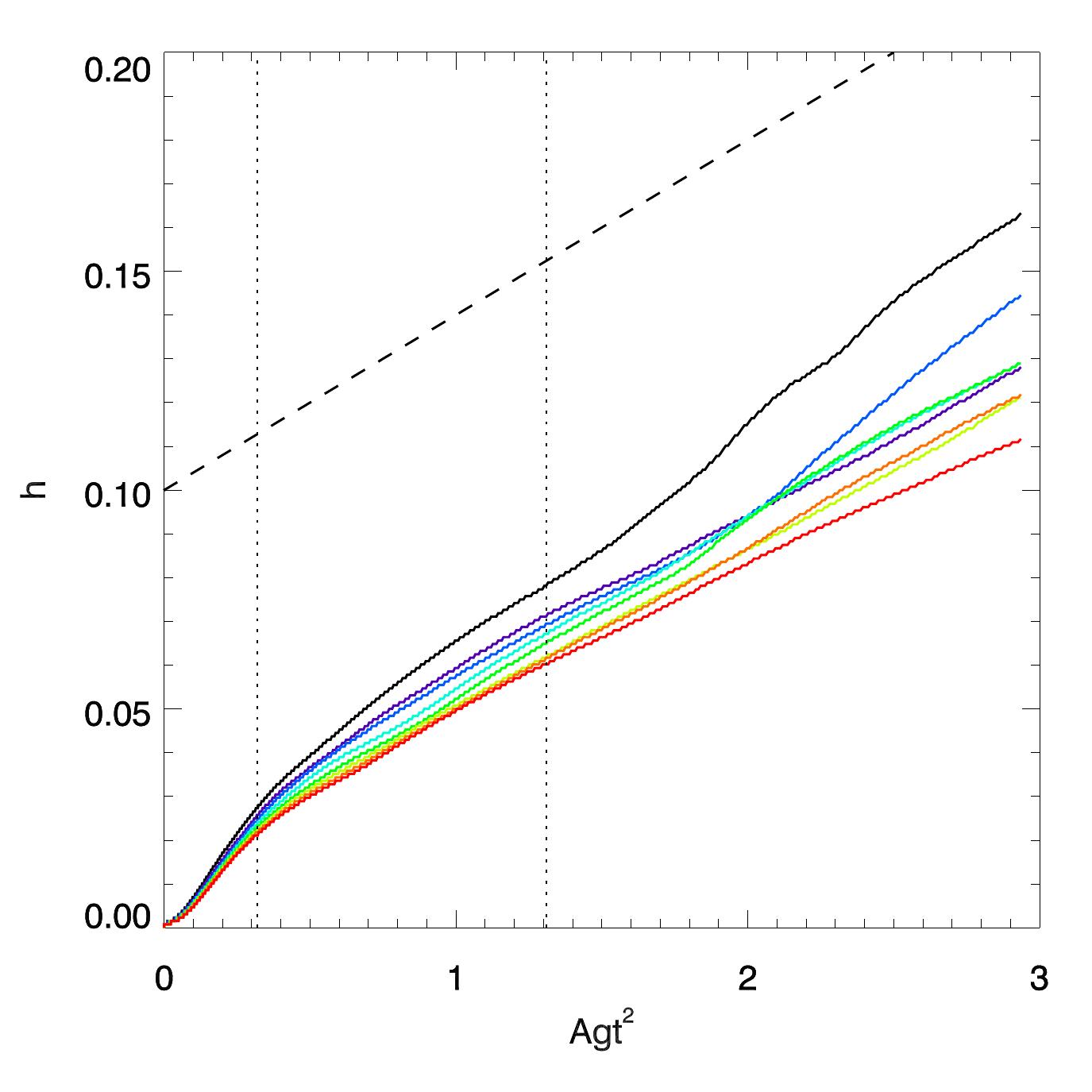}
\includegraphics[scale=0.4]{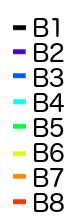}
\caption{Plots showing the development of bubble height as a function of $Agt^2$ for the B simulations (increasing magnetic field strength). Dotted lines mark $t~=~2$ and $t~=~4$; simulations end at $t~=~6$. Dashed line shows the slope of $\alpha~=~0.04$.}
\label{maggraph}
\end{figure}

The measure of bubble height $h$ is taken as being the highest point at which the average density over the x-y-plane is $\langle\rho_{z}\rangle~\leq~9.5$, which should return a position at the average height of all bubbles; Figures~\ref{sims1}~and~\ref{sims2} show the sharp density gradients at the edges of the mixing region. Figure~\ref{maggraph} shows the development of the highest $\langle\rho_{z}\rangle~=~9.5$ (\emph{i.e.}, equation~\ref{rteqnl2}) for each simulation in the magnetic field varying set. The gradient of the slopes defines the relative rate at which bubbles grow, and it is apparent that this decreases across the simulations from B1 to B7 from visual inspection, suggesting that increased magnetic field strength will yield a reduced (non-linear) growth rate (as well as agreeing with the analytic prediction that linear growth rate decreases with magnetic field strength; see equation~\ref{rtgrow2}). 

The early linear phase of the RTI can be seen in Figure~\ref{maggraph}, characterised by a rapid growth. The rate of growth appears to then suddenly decrease at the same point in all simulations, continuing thereon with a relatively steady dependence on $t^2$. Some of the B simulations diverge from this dependence towards later times, but this is likely to be the result of the formation of large, coherent flows developing which weaken the statistics of the averaging process.

\begin{figure}
\centering
\includegraphics[scale=0.55]{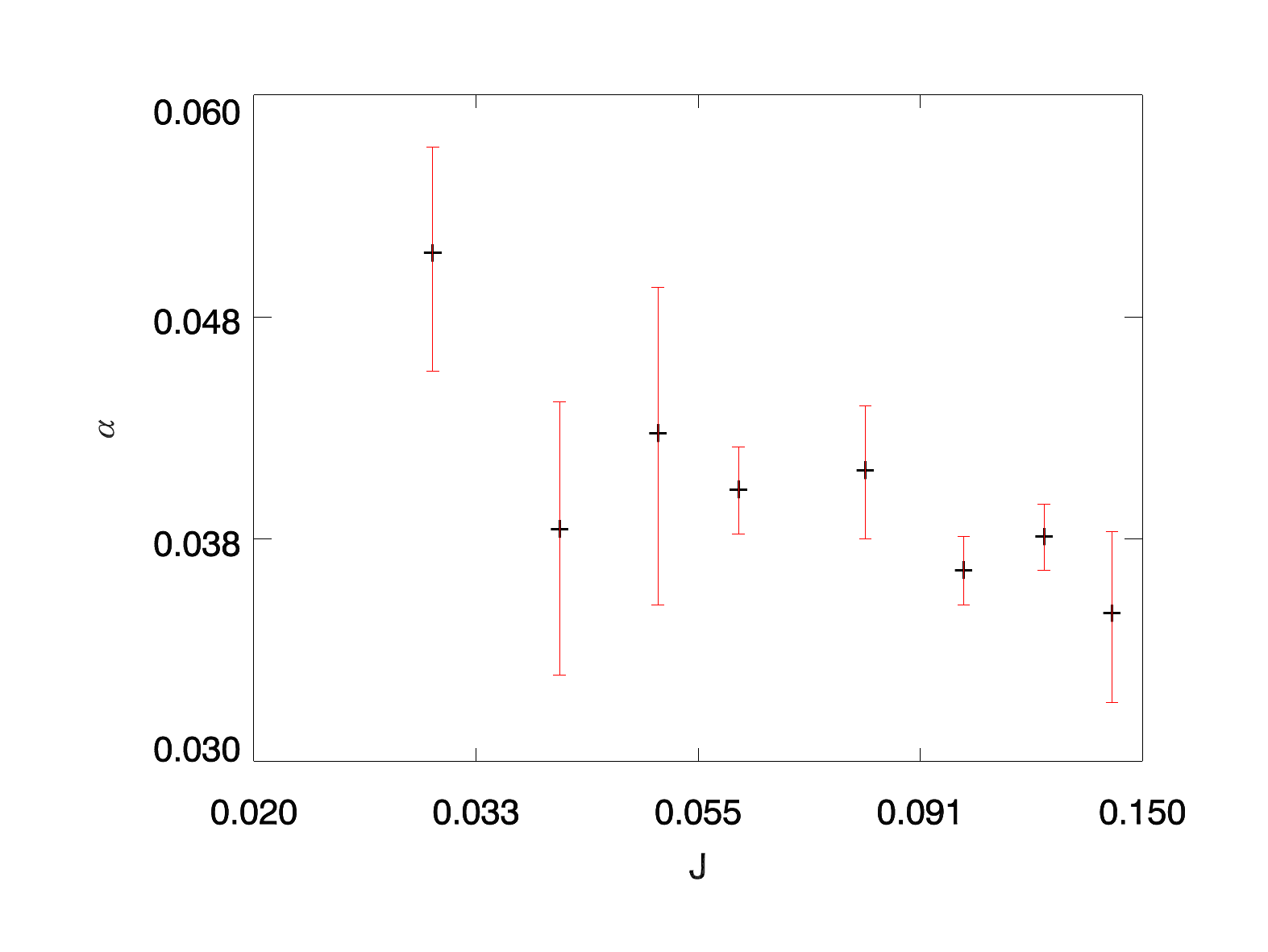}
\caption{$\alpha$ plotted against $J$ for all B simulations - note the scaling of the axes is logarithmic.}
\label{a-j}
\end{figure}

The non-linear growth rate is calculated by finding the rate of change of $h$ relative to $Agt^2$ for each time-step (by fitting a linear regression to the surrounding 100 datapoints) in the non-linear regime, and taking the mean value. This gives a value for $\alpha$, as well as the standard deviation, $\sigma$, for each simulation. These are given in Table~\ref{simtab}, and are plotted against $J$ for all B simulations in Figure~\ref{a-j}, with error bars representing the standard deviation. 

\begin{figure}
\centering
\includegraphics[scale=0.55,trim={2cm 0 0 0}]{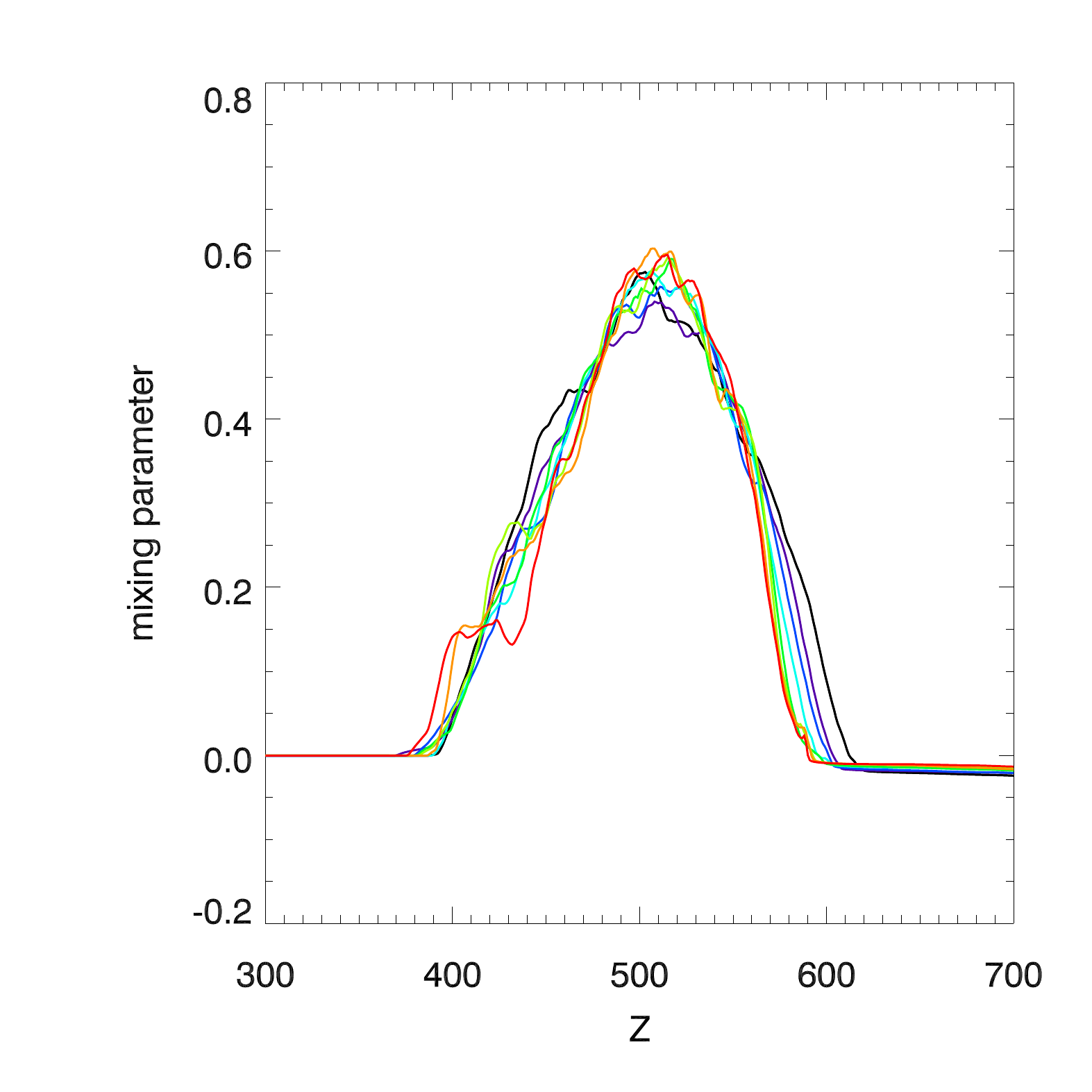}
\includegraphics[scale=0.4]{magkey.png}
\caption{The average degree of mixing over the height of all B simulations at $t~=~4$.}
\label{mixb}
\end{figure}

A mixing parameter may also be defined as $\Theta~=~4 \langle f_h f_l \rangle$ \citep{stone_magnetic_2007}, where $f$ denotes the fractional amount of upper and lower density material, respectively. That is $f_{\rm l}~=~|10~-~\rho|/9$ and $f_{\rm h}~=~|1~-~\rho|/9$, where these are averaged over the $x-y$-plane. These data are plotted for all B simulations at $t~=~4$ in Figures~\ref{mixb}. 

\begin{figure}
\centering
\includegraphics[scale=0.55]{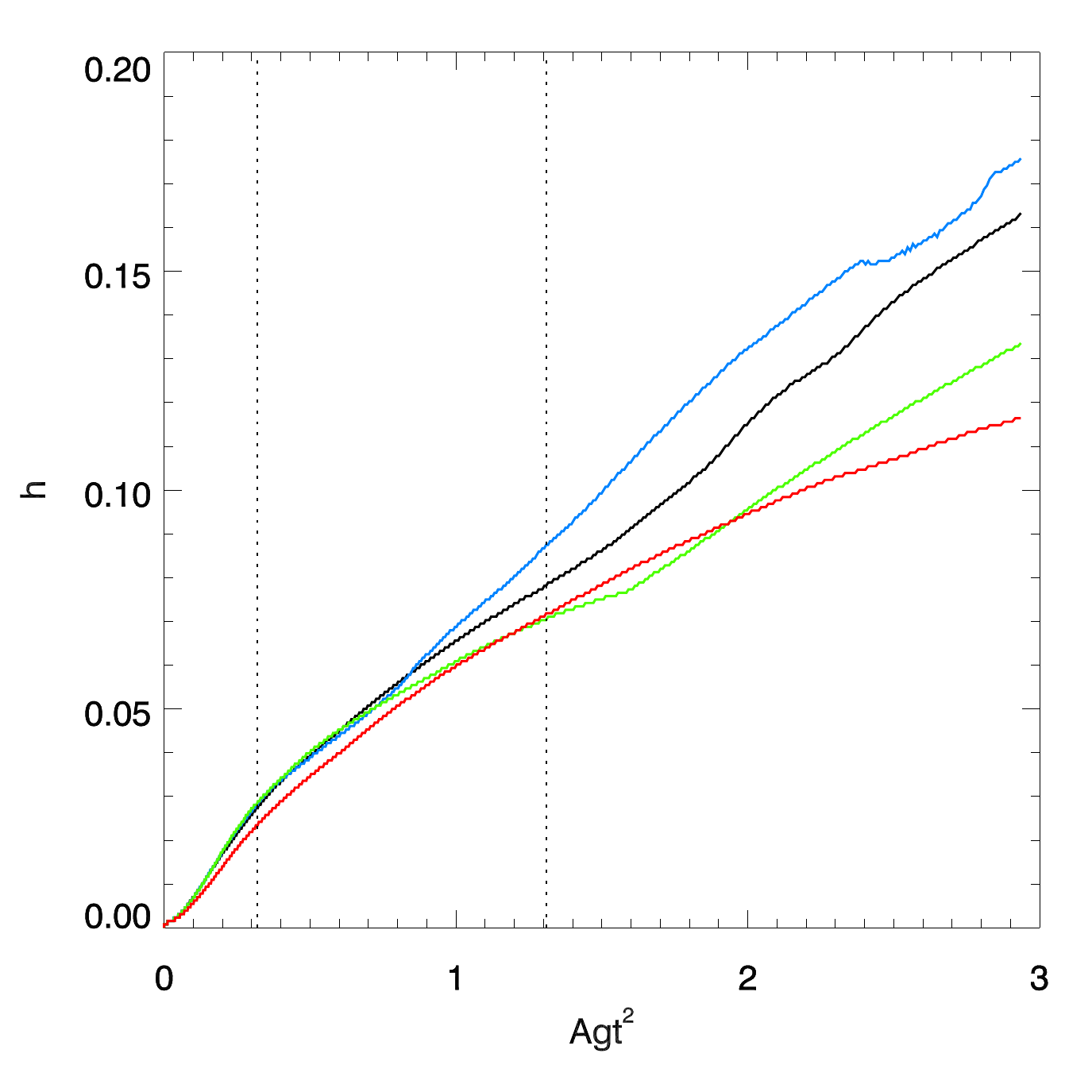}
\includegraphics[scale=0.4]{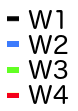}
\caption{Plots showing the development of bubble height as a function of $Agt^2$ for the W simulations (decreasing domain width). Dashed lines mark $t~=~2$ and $t~=~4$; simulations end at $t~=~6$.}
\label{widgraph}
\end{figure}

Similar plots of the W simulations are shown in Figure~\ref{widgraph}, which also appear to show a change in $\alpha$ across the simulations, listed in Table~\ref{simtab}, where $L/\lambda_u$ is the only parameter which is decreasing between simulations. This suggests that magnetic field is not necessarily having a strong effect on the non-linear growth of the RTI, though the trend in $\alpha$ is much clearer and more correlative in the initial (B) set. $\alpha$ is also found for the W simulations over the non-linear phase;  these values are listed in Table~\ref{simtab}. 

\begin{figure*}
\centering
\includegraphics[scale=0.4]{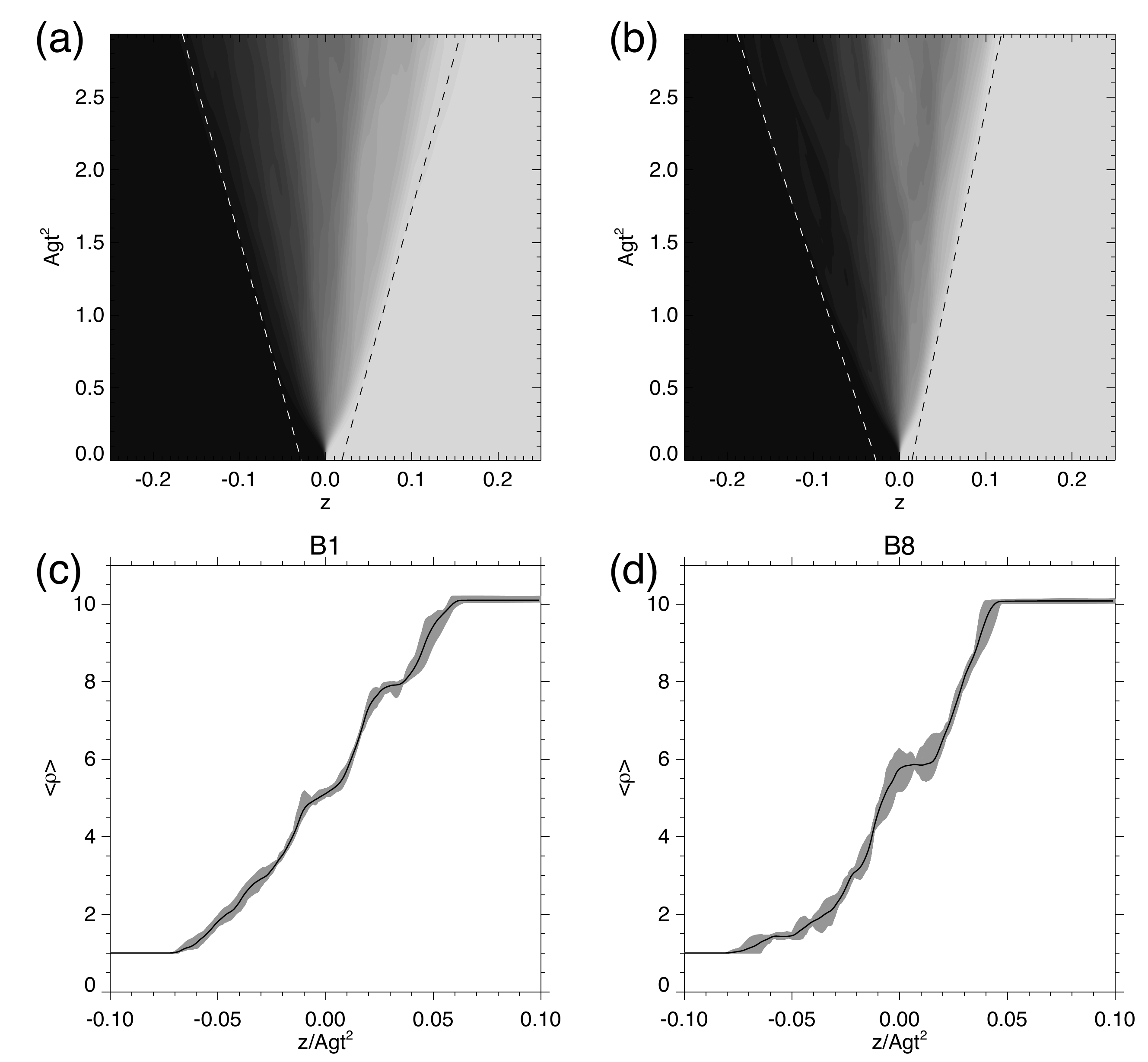}
\includegraphics[scale=0.4]{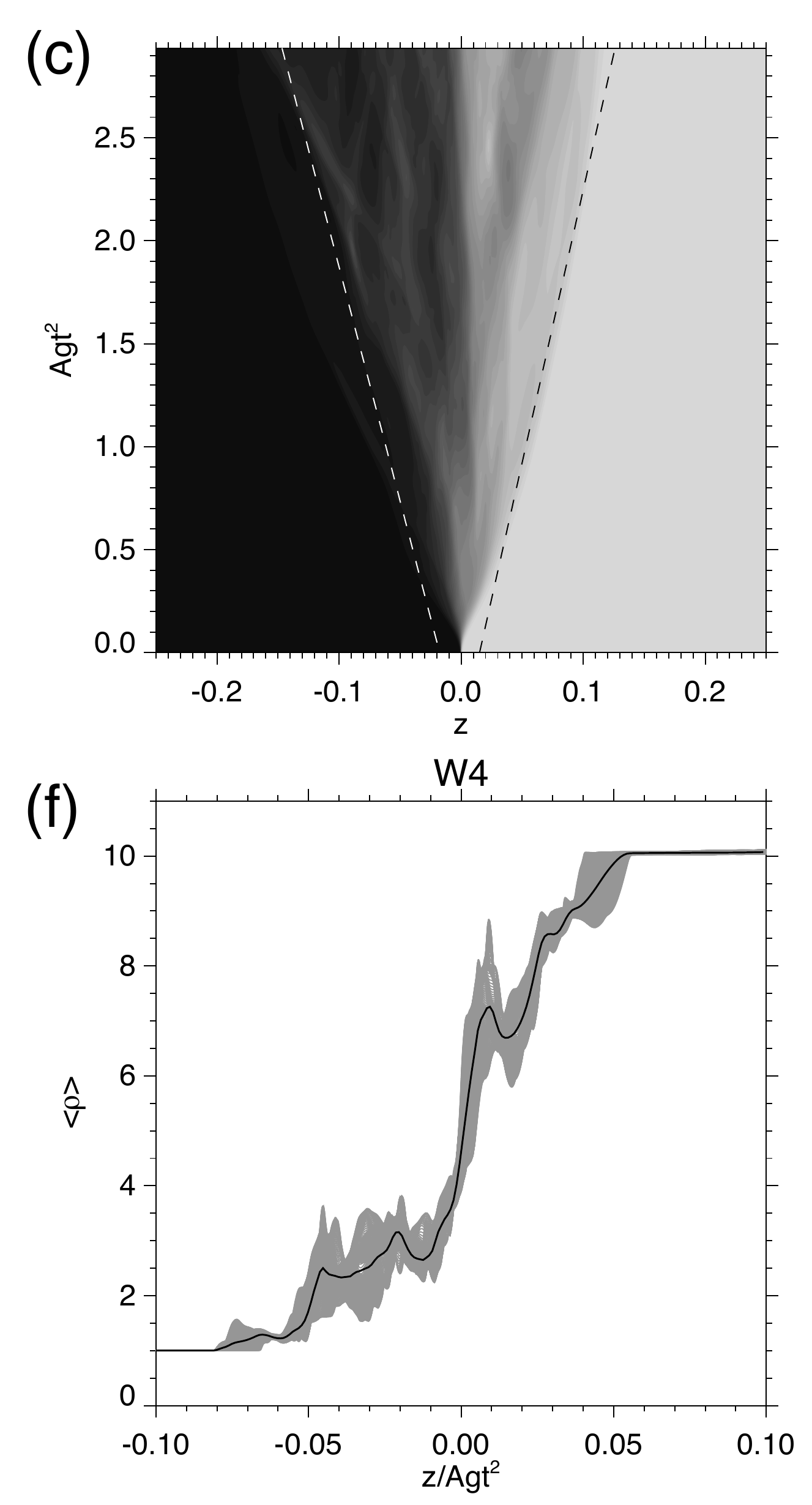}
\caption{\textbf{Panel (a) shows a contour plot of the horizontal mean of the density for B1, (b) is the same for B8, and (c) is the same for W4. The dashed white and black lines signify $\alpha=-0.047$ and $\alpha=0.047$, respectively, for panel (a), $\alpha=-0.055$ and $\alpha=0.0354$ for panel (b), and $\alpha=-0.044$ and $\alpha=0.038$ for panel (c).
 Panel (d) shows the mean density distribution plotted against $z/Agt^2$ for the final 200 snapshots of the simulation for B1, (e) is the same but for B8, and (f) is the same but for W4. Note that the values that would be calculated for $\alpha$ from this figure are over estimates due to the initial expansion of the system.}}
\label{mean_den}
\end{figure*}

Figure \ref{mean_den} shows the expansion of the mixing layer for simulations B1 and B8. The top three panels are contour plots of the $\langle\rho\rangle$ for B1, B8 and W4 respectively with $z$ as the horizontal axis and $Agt^2$ for the vertical. The dashed white and black lines highlight the lower and upper limits of the expansion of the mixing with $t^2$ dependence. The values for $\alpha$ associated with this are $-0.047$ and $0.047$ for B1, $-0.055$ and $0.0354$ for B8, and $\alpha=-0.044$ and $\alpha=0.0378$ for W4. The bottom three panels of this figure show the average of the mean density profiles over the final 200 snapshots of the simulation for B1 (panel d), B8 (panel e) and W4 (panel f). Note that the values for $\alpha$ that can be calculated from this plot are overestimates due to the fast initial expansion of the layer, this issue is discussed in \cite{cabot_reynolds_2006}.

\section{Discussion} \label{disc}

Simulations of RT-unstable plasma (governed by an ideal equation-of-state) are conducted close to the incompressible limit, with homogenous magnetic fields aligned perpendicular to gravity (parallel to the density-jump interface in the initial conditions), and the impact of the strength of the embedded magnetic fields on the development of the non-linear regime of the RTI is examined. From visual inspection of Figures~\ref{maggraph} and~\ref{a-j}, and from the calculated $\alpha$ in Table~\ref{simtab}, it is apparent that increased magnetic field strength leads to reduced non-linear growth rate. However, it is not possible to draw a linear regression trend through all data in Figure~\ref{a-j}; that is, the decrease in non-linear growth rate as magnetic field strength increases is not smooth. Figure~\ref{a-j} (as well as Table~\ref{simtab}) shows, for example, that the growth rate for the B2 simulation is much lower than that of B3 and even B4, each of which have progressively stronger magnetic fields.

Some Values of $\alpha$ returned in this work are not necessarily equal to that of other studies (in fact a single value of $\alpha$ is not agreed upon between very many studies). This may be explained by \cite{glimm_critical_2001}, who show that numerical dissipation effects (such as mass diffusion and viscosity) due to algorithmic differences and differences in simulation duration are the main reasons for this discrepancy across studies, whilst within this study these effects are constant across simulations and so the calculated non-linear growth rates are suitable to be compared to one another, though comparisons of absolute values with other work is less trustworthy.

The reduced growth rate for stronger fields could be due to an increased magnetic tension ($=(\mathbf{B}\cdot\nabla)\mathbf{B}/\mu_0$) in stronger magnetic fields. Greater magnetic tension reduces the free energy of the system along the direction of the magnetic field line, suppressing the modes of the instability in this dimension, and with the reduced contribution to the whole system from these modes, the overall growth-rate will be reduced. This is not in disagreement with \cite{stone_magnetic_2007}, who conclude that the RTI non-linear growth rate is faster when a magnetic field is added relative to the hydrodynamic case, explained by the reduced interface-mixing from suppressed Kelvin-Helmholtz roll-ups by magnetic tension.

Whilst magnetic field strength is the only parameter altered between successive runs of the initial B set, this leads to a secondary constraint on the physics of the system: the ratio of the dominant scale of the instability ($\lambda_u$, from equation~\ref{rtgrow3}) to domain width ($L_x$) may constrain the bubbles which develop along the direction of magnetic field. If this is below unity, equation~\ref{rtgrow3} gives a characteristic scale size of the simulation as larger than the simulated domain. Moreover, as explained in Section~\ref{intro}, in order for equation~\ref{rteqnl2} to apply to the system (\emph{i.e.} the equation relating $h$ and $\alpha$), we must fulfil $L~>~h~>~\lambda_{max}$. In order to investigate whether $\lambda_u/L$ affects the growth rate (either linear or non-linear) RTI, as well as to examine the behaviour of a system which violates the $L~>~h~>~\lambda_{max}$ constraint, a further set of simulations were run with constant magnetic field strength but variable width. These displayed non-linear growth rates which showed similar variation to the first B set, but also with no apparent correlation and greater uncertainty. This is highlighted by comparing Figures~\ref{maggraph} and~\ref{widgraph}, as well as the standard deviations in Table~\ref{simtab}. The height vs. time-squared plots are visually more correlative for the B set of simulations than the W set, and the standard deviation on $\alpha$ is somewhat higher for the W simulations (see Table~\ref{simtab}. The lack of correlation in $\alpha$ for the W simulations gives validity to the results of the B simulations which do appear to display a trend, indicating that enhanced magnetic field strength leads to reduced non-linear growth-rates.

Measured non-linear growth rate $\alpha$ for the W simulations also highlighted the importance of the $L~>~h~>~\lambda_{max}$ requirement. It is apparent that where this is violated (W4) the RTI does not develop in the same manner as all other simulations conducted - from visual examination of Figure~\ref{widgraph}, the slope of W4 mixing region height does not follow the same trend as other simulations plotted in this figure, nor in any from Figure~\ref{maggraph}. This can also be seen in Table~\ref{simtab}, which shows the standard deviation on W4's calculated $\alpha$ is the highest of all simulations.

There are further implications for the 3D RTI with unidirectional embedded magnetic fields; if the magnetic tension opposes the deformation of the field lines induced by plasma motions, but only in the direction of the field lines, this is essentially reducing the free energy of the system in only this dimension. It would therefore follow that by introducing progressively stronger magnetic fields, the system approaches a quasi-2D domain, and hence the growth rate of the instability would be reduced. This is demonstrated by \cite{kane_two-dimensional_2000}, who show that $\alpha$ is reduced by roughly 30\% in the 2D case relative to the 3D case. This can explain the result obtained here for reduced $\alpha$ with increased magnetic field strength.

In order to measure the non-linear growth rate of the RTI, the gradient of a curve such as those plotted in Figure~\ref{maggraph} is often commonly used. However, a precise value of $\alpha$ for each simulation is difficult to measure, as this is an attempt to describe the average behaviour of the non-linear system; the non-linearity itself implies fluctuations which will change $\alpha$ on small timescales. For this reason, the mean value of the rate of change of the height of the mixing region over the non-linear regime was used to calculate alpha.

Furthermore, the transition between linear and non-linear regimes of the instability is ill-defined, so the $\alpha$ measurement starting point can be difficult to choose. As a good approximation, the eigenfunction for the vertical velocity $v_z$ is given by
\begin{equation}
v_z(z)~=~A e^{-k|z|}
\end{equation}
\citep{chandrasekhar_hydrodynamic_1961}, which implies that $1/k$ can be used as the vertical scale through which the perturbation can travel before it reaches its non-linear saturation. Figures~\ref{sims1} and~\ref{sims2} indicate that this is achieved just after $t~=~2$ for the low-strength magnetic field cases, and at later times as this is increased. It is possible that non-linear saturation has not been reached for the strongest magnetic fields by the end of the simulation; the bottom-right of Figure~\ref{sims1} suggests that the height of the mixing region has not yet reached the observed wavelength. However, from Figure~\ref{maggraph}, B8 indeed displays the shift of behaviour into an apparent $t^2$ dependence at roughly the same time as all other simulations, suggesting that a vertical scale which depends on the magnetic field strength (indirectly, \emph{i.e.}, through modification of $1/k$) may not be the best method of identifying non-linear saturation. Note it has been shown that the the initial expansion of the system when dominated by linear modes can also display the same $t^2$ dependence \citep{hillier_nature_2016}, which is consistent with our findings in this paper.

The results for the mixing parameter in Figure~\ref{mixb} show that increasing the magnetic field does not seem to drastically increase the density mixing of the system, which is in contrast to the results found for including shear in the magnetic field of \cite{stone_magnetic_2007}. 

The non-linear growth rate is also calculated from the upper and lower limits of the expansion of the mixing layer (see Figure~\ref{mean_den}). This finds that for stronger magnetic field cases, there is an asymmetry between the upwards growth of the bubbles and the downwards growth of the spikes. The difference between the two $\alpha$ values is approximately 0.09 for both B1 and B8, so though B1 grows quicker upwards, the overall expansion of the layers in both simulations is approximately the same. W4, however, has a smaller difference of approximately $0.08$ which we can attribute to the simulation becoming more 2D like which is know to reduce the value of $\alpha$.
This leads to the conclusion that the development of asymmetry from B1 to B8 is to some extent a magnetic effect and not purely as a result of the stronger magnetic field rendering the B8 simulation more 2D like. A note of caution must be presented with this, because W4 by necessity has poorer statistics for the averaging (due to less structure being averaged across resulting in individual plumes carrying greater weight in the averaging process), and so it is harder to draw strong conclusions.

From panels (d) and (e) in Figure~\ref{mean_den}, it can be seen that the asymmetry in the expansion of the mixing layer of B8 also manifests itself in the distribution of the mean density of the mixing layer. For B1, the density profile is approximately linear, but for B8 a long tail exists in the distribution at the lower level. The cause of this asymmetry is unclear, however, it does suggest that the reduced growth rate for enhanced magnetic field strengths may not be an accurate interpretation of these results. This asymmetry reflects a reduction of the erosion of the upper dense layer (as evidenced by the lower $\alpha$ value). As the erosion of this layer reflects the release of the gravitational potential energy that is driving the mixing, it is no surprise that on calculating the energy release, given by
\begin{equation}
E_{GP}=\int_{-0.1}^{0.1} (\langle \rho(z') \rangle - \rho_{\rm init}(z'))g z' dz'
\end{equation}
where $z'=z/Agt^2$ and $\rho_{\rm init}$ is the initial density distribution, the energy released in simulation B8 is only 76\% of that of B1. 
As energy is a conserved quantity, a reduction in the amount of gravitational energy released must necessarily result in reductions in the amount of energy found in the turbulent components of the velocity and magnetic field, because there is less energy released to drive the growth of these components.
Therefore, the fact that the increase in magnetic field strength reduces the gravitational energy release leads to a reduction in the turbulence driven by this instability.

The effect of $J$ (see equation~\ref{jackpam}) is explored over a parameter space ranging from the highest value conducted in previous work \citep{stone_magnetic_2007}, and a maximum value which is an accurate representation of plasma observed to be RT unstable in the crab nebula \citep{hester_wfpc2_1996}, and is factor two lower than that predicted for erupted solar filament plasma such as the study by \cite{innes_break_2012}. The simulations are conducted on plasma at the incompressible limit, and in both of these examples mentioned, the sound speed is believed to be greater than velocities of any fluid motions. Therefore the results of this study are applicable to the observations of astrophysical plasmas which initially prompted this investigation. In the initial conditions of the simulations, a homogenous magnetic field parallel to the interface is used. Such a configuration seems unlikely to exist in nature, however, it can be very difficult to observe the magnetic structure within astrophysical plasmas. Since no previous studies have investigated the effect of the strength of the magnetic field on the non-linear RTI, this is the most straightforward case to begin investigating. The results demonstrate that the strength of magnetic fields embedded in plasmas will affect the development of the non-linear mixing by the RTI.

\section{Conclusion} \label{conc}

Simulations of the RTI were conducted in order to better understand how magnetic field strength may affect the growth rate of this instability. It has been found from previous work that non-linear growth rate is enhanced when a strong magnetic field is present (c.f. the hydrodynamic case), however, this study has found that increasing the strength of the magnetic field leads to a decrease in non-linear growth rate of the rising bubbles of the instability. This is speculated to be due to higher magnetic tension requiring greater energy in order for the frozen-in plasma to move. Non-linear growth rates were found to converge on $\sim$0.039 for the strongest magnetic fields studied, however, the decrease in non-linear growth rate in the results presented in this work is by no means a smooth one; for instance, the growth rate of the second-weakest field is almost equal to that of the second-strongest field.

Since altering magnetic field strength in a fixed domain has the indirect effect of changing the ratio between domain size and dominant mode (\emph{i.e.} characteristic length-scale) of the RTI, this was investigated for a set of simulations with constant magnetic field strength. Whilst the calculated growth rates for these simulations also showed some difference, the results were not correlative at all, as the first set was, implying that the interpretation of stronger magnetic fields reducing the non-linear development of the RTI is accurate.

Finally, the non-linear growth rate was estimated in a different way from both the rising bubbles and the falling spikes; these results corroborated the earlier conclusion that stronger fields yield lower growth rates in the rising bubbles, however, this also indicated that the growth rate of the falling bubbles is enhanced by approximately the same order. This suggests that stronger fields do not enhance the development of the RTI, but in fact create an asymmetry. This leaves the $\alpha$ value as a measure of the rate at which the dense layer is eroded, and its energy is released. Where we find that stronger magnetic fields slow down the release of this energy.

Whilst the simulations may not perfectly describe observed RT-unstable plasma throughout the universe ({\emph e.g.}, the simulations are conducted approximately in the incompressible limit, with homogenous magnetic fields, under the ideal gas equation), they are conducted in a parameter space relevant to the relative effects of magnetic tension and gravity on such astrophysical plasmas. Hence, we have shown that increased magnetic field strength reduces the non-linear development of the RTI in situations where the instability is observed throughout the universe. This arises from the magnetic tension which acts against the deformation of field lines, and hence acting against the instability in the dimension aligned with the field. By reducing the growth in this way ({\emph i.e.}, in one dimension), the development of the instability overall in the system is reduced.

\begin{acknowledgements}The authors wish to thank Davina Innes for inspiring this project, as well as invaluable discussions and advice; and thanks to LiJia Guo for assistance with performing the numerical experiments.
Jack Carlyle conducted this work during a PhD joint-funded by University College London and The Max Planck Institute for Solar System Research, as well as whilst receiving funding as a Research Fellow from the European Space Agency. 
Andrew Hillier is supported by his STFC Emest Rutherford Fellowship grant number ST/L00397X/2.
\end{acknowledgements}

\end{document}